\documentclass[aps,twocolumn,showpacs,preprintnumbers,amsmath,amssymb,prl]{revtex4}

\usepackage{graphicx}% Include figure files
\usepackage{color}

\usepackage{dcolumn} % Align table columns on decimal point
\usepackage{bm} % bold math

\begin{document}

\title{Observation of resonance-assisted dynamical tunneling in an asymmetric microcavity}
\author{Hojeong Kwak}
\affiliation{School of Physics and Astronomy, Seoul National University, Seoul 151-742, Korea}
\author{Younghoon Shin}
\affiliation{School of Physics and Astronomy, Seoul National University, Seoul 151-742, Korea}
\author{Songky Moon}
\affiliation{School of Physics and Astronomy, Seoul National University, Seoul 151-742, Korea}
\author{Kyungwon An}
\email{kwan@phya.snu.ac.kr}
\affiliation{School of Physics and Astronomy, Seoul National University, Seoul 151-742, Korea}

\date{\today}

\begin{abstract}
We report the first experimental observation of the resonance-assisted dynamical tunneling (RADT) in the inter-mode interaction in an asymmetric-deformed microcavity.
A selection rule for strong inter-mode coupling induced by RADT was observed on angular mode numbers as predicted by the RADT theory. 
In addition, the coupling strength was measured to be proportional to the square of the phase-space area associated with the nonlinear resonance involved in RADT. 
The proportionality constant was found to depend only on the nonlinear resonance, supporting the semiclassical nature of RADT. 
\end{abstract}


%03.65.Sq, Semiclassical theories in quantum mechanics    
%42.55.Sa, Microcavity and microdisk lasers    
%03.65.Xp. Traversal time (quantum mechanics)
%05.45.Mt, Quantum chaos
	
\maketitle

%\section{Introduction}
Dynamical tunneling is a quantum-mechanical tunneling phenomenon to occur between dynamically separated classical trajectories \cite{Davis81}. 
Dynamical tunneling is known to be affected by the underlying classical phase-space structure as extensively studied in chaos-assisted tunneling \cite{Lin90,Bohigas93,Tomsovic94,Steck01, Hensinger01,Dembowski00}.
It is also predicted that the presence of nonlinear resonance structure can enhance dynamical tunneling.

In an integrable multi-dimensional system, classical trajectories appear as invariant tori on the Poincar\`e surface of section (PSOS). 
The phase-space projections of quantum eigenstates are then localized along these tori. 
In the presence of perturbation, invariant tori are deformed following the Kolmogorov-Arnold-Moser (KAM) scenario
and some orbits evolve into a {\em chain}-like nonlinear resonance structure qualitatively distinguished from the KAM tori. 
The nonlinear resonance structure can then strongly enhance a tunneling process between the modes localized along nearby invariant tori when specific conditions are satisfied. This type of enhanced dynamical tunneling is theoretically known as the resonance-assisted dynamical tunneling (RADT) \cite{Brodier,Almeida84}. 
 
RADT is a universal phenomenon expected to occur in any weak-perturbed systems of near-integrable or mixed phase space since the theory of RADT does not depend on the details of the Hamiltonian. 
RADT has thus been theoretically studied in various systems such as periodic-driven pendula \cite{Mouchet06},  Rydberg atoms under periodic perturbation \cite{Wimberger06}, quantum accelerator modes \cite{Sheinman06} and multi-dimensional molecules \cite{Keshavamurthy05,Keshavamurthy05r}. 
RADT has also been extensively studied in one-dimensional time periodic quantum maps such as the kicked Harper model and the kicked rotor \cite{Brodier, Eltschka05, Loeck10, Wisniacki11}.
However, to the best of our knowledge, 
there exist no experiments yet confirming the prediction of the RADT theory.
 
In this Letter, we report the first experimental observation of RADT between modes in an asymmetric microcavity. 
Strong RADT and thus large avoided crossing (AC) of energy levels were observed between unperturbed-basis modes when their angular mode numbers differ by an integer multiple of the number of islands in the associated nonlinear resonance chain in the phase space. The AC energy gap, approximately twice of the coupling strength by RADT, increased in proportion to the square of the phase-space area associated with the nonlinear resonance chain as the cavity boundary deformation increased. 
Moreover, the proportionality constant was dependent only on the nonlinear resonance.
These observations are the key predictions of the RADT theory. 
 
%\section{Strong and Weak Interaction}
The specific physical system we consider is a two-dimensional asymmetric-deformed microcavity made of a liquid jet column of ethanol (refractive index $n$=1.357) doped with laser dye styryl (LDS) molecules as fluorophore. The details of our liquid jet system are described elsewhere \cite{Yang06}. In short, the cavity boundary shape is approximately a quadru-octapole given by $r(\phi)\simeq
a(1+\eta\cos2\phi+\epsilon\eta^2\cos4\phi)$, where $a\simeq (15.1 \pm0.1) \mu$m, the mean radius of the cavity, and $\epsilon=0.42\pm0.05$ \cite{Moon08}. The deformation parameter $\eta$ can be continuously tuned from 0 to 26\% by changing the jet ejection pressure.

Before performing experiments, 
we surveyed the interactions among unperturbed modes in our system in numerical simulations. We employed the boundary element method \cite{Wiersig03} and calculated the quasi-eigenvalues and associated Husimi functions for the same size and shape as our liquid-jet microcavity.
The quasi-eigenvalues are presented in terms of the size parameter $ka$ with $k=2\pi /\lambda$ the wavevector. 

\begin{figure}
\includegraphics[width=3.4in]{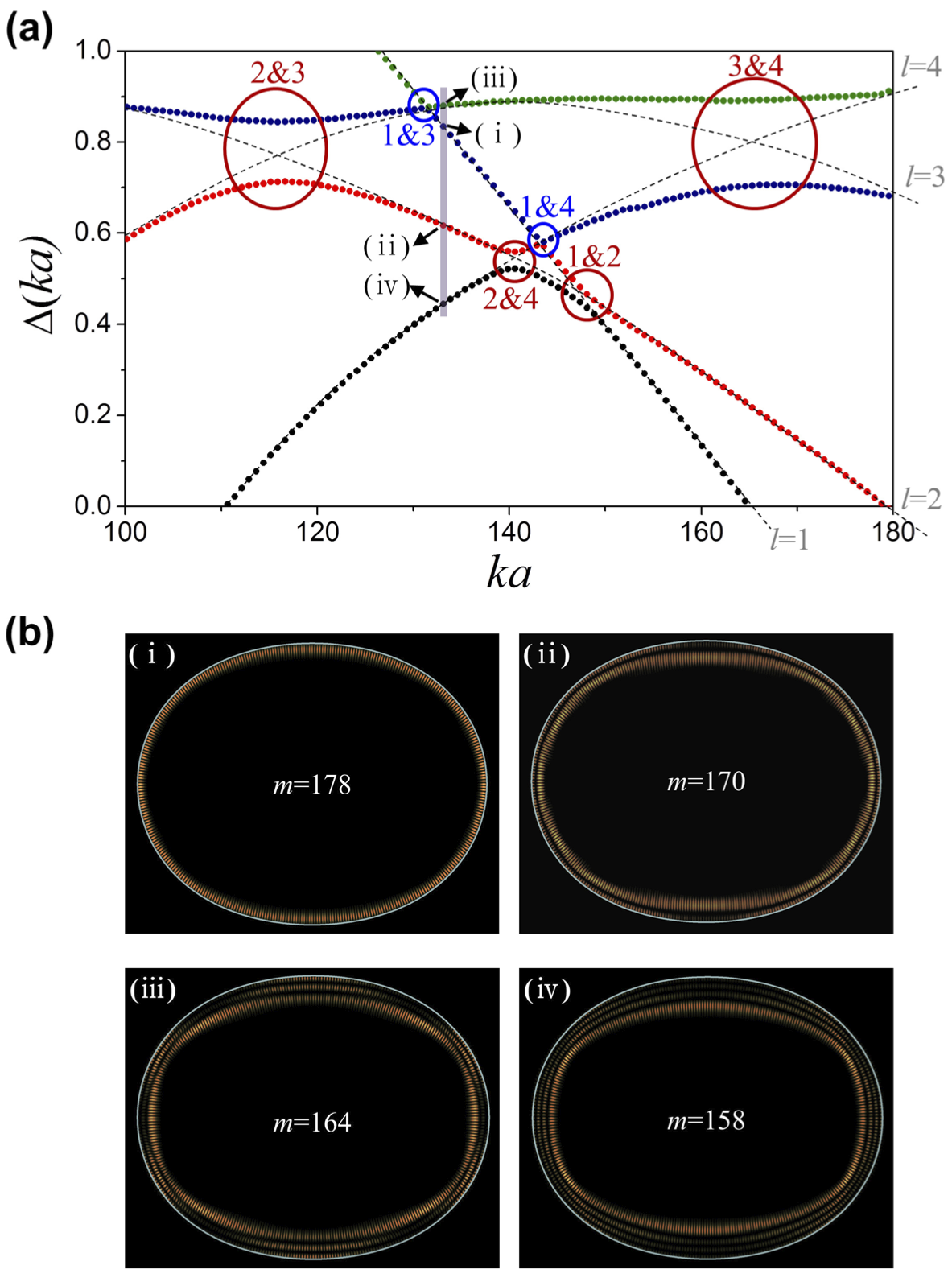}
\caption{ (a) Mode dynamics diagram showing relative frequencies $\Delta(ka)$ of $l$=1, 2, 3 and 4 modes calculated with respect to a reference frequency in the range from $ka\sim$100 to 180 when $\eta$=0.16.  (b) Spatial mode-distribution intensity plots of the $l$=1, 2, 3 and 4 modes marked by arrows in (a). These modes are far from the interaction regions. The solid line indicates the cavity boundary.} 
\label{Spectrum}
\end{figure}

Figure \ref{Spectrum}(a) shows inter-mode dynamics when $\eta$ = 0.16. For this, we first numerically find high-$Q$ mode spectra in the range from $ka \simeq $100 to 180 and identify {\em uncoupled} mode groups labeled by {\em radial}  mode order $l$ (=1, 2, 3, 4) in the increasing order of their free spectral ranges in the {\em uncoupled} region. 
We then define a sequence of reference frequencies of a regular spacing and measure the relative frequencies $\Delta(ka)$ of each mode group with respect to the reference frequencies.
The relative frequencies of all four mode groups are plotted in the mode dynamics diagram in Fig.\ \ref{Spectrum}. 
Detailed information on the uncoupled mode labeling and the relative frequency measurement is described elsewhere \cite{Lee09}.

Each mode group in Fig.\ \ref{Spectrum} more or less follows a {\em diabatic} line unless it encounters other mode groups. 
Diabatic lines are shown as dashed lines with associated $l$ values denoted in Fig.\ \ref{Spectrum}(a).
When mode groups encounter each other, they exhibit AC's.
The AC gap -- defined as the smallest energy separation of two interacting levels or mode groups -- is approximately twice the coupling strength between them (see below for more explanation). By inspecting the AC gap, we can qualitatively identify two types of interactions, {\em strong} (circled red) vs.\ {\em weak} (circled blue) interactions. 

\begin{figure}
\includegraphics[width=3.4in]{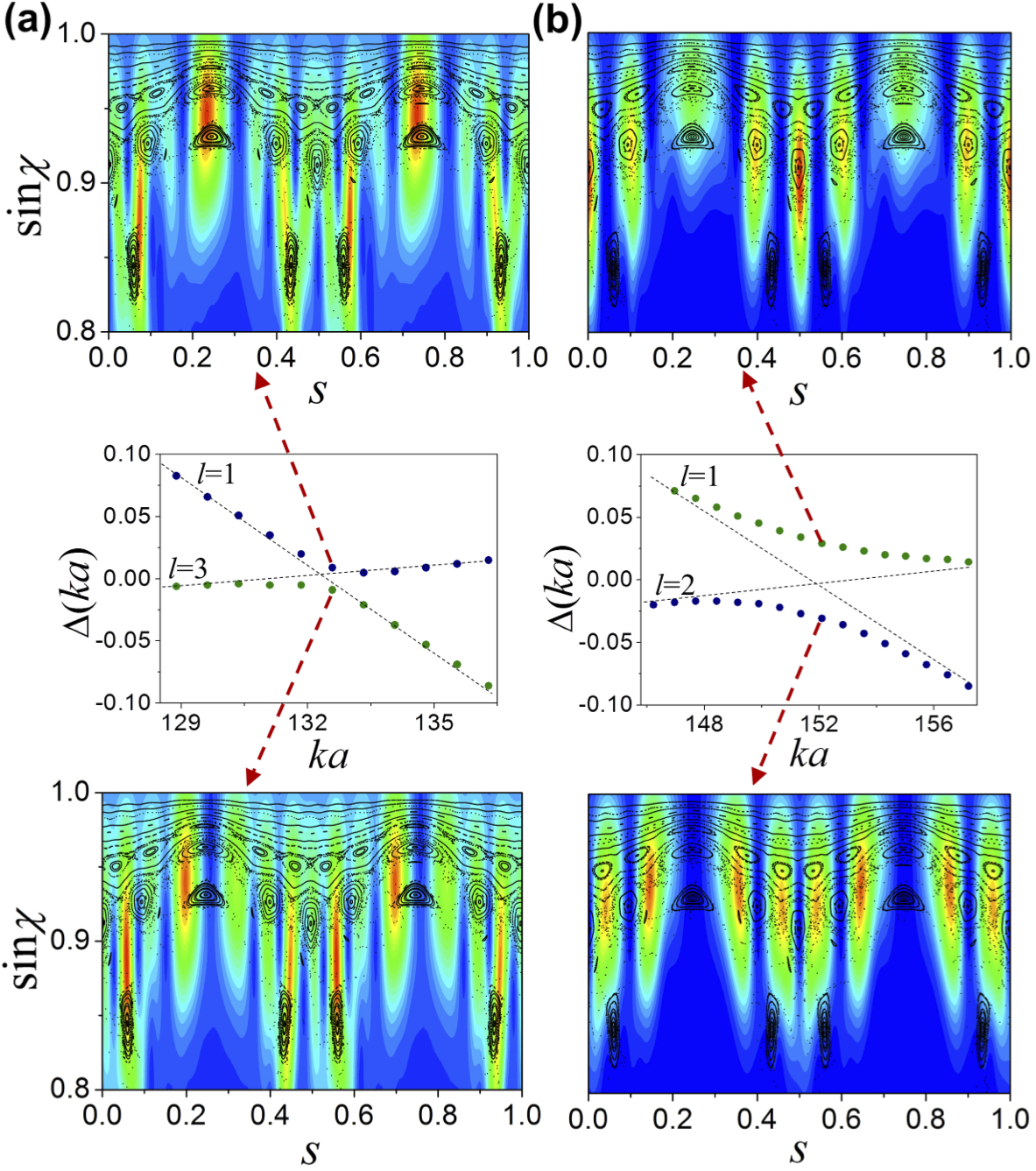}
\caption{(a) Relative frequencies of $l$=1 and 3 modes near the AC region and the Husimi plots of the two modes at the closest encounter, marked by dashed arrows, when $\eta=0.19$. (b) The same plots for $l$=1 and 2 modes when $\eta=0.19$.}
\label{Husimi}
\end{figure}

To further investigate the distinction between these two types of interactions, 
we consider the Husimi functions of the quasi-eigenmodes near the interaction region and compare them with the Poincar\'e surface of section (PSOS) of the augmented ray dynamics \cite{Unterhinninghofen08} as in Fig.\  \ref{Husimi}. 
The PSOS is presented in the Birkhoff's coordinates ($s, \sin \chi$), where a ray is reflected off the cavity boundary at the normalized arc-length coordinate $s$ $(0\le s \le 1)$ along the boundary from the major axis with an incidence angle $\chi$.
When two modes experience a weak interaction as in Fig.\ \ref{Husimi}(a), 
the Husimi functions upon the closest encounter exhibit mixing of their Husimi functions   
that we would have far away the avoided-crossing region following the diabatic lines.
This feature is what we generally expect in AC of quasi-eigenmodes.

On the other hand, when two modes undergo a strong interaction as shown in Fig.\ \ref{Husimi}(b),
their Husimi functions are well localized 
along the stable or unstable periodic orbits associated with a classical nonlinear resonance structure in the phase space. 
All of the strong and weak interactions in Fig.\ \ref{Spectrum}(a) show the above tendencies of the Husimi functions, respectively, upon AC's. 
The Husimi function localization phenomena, in particular, suggests that the strong interactions must be related to the nonlinear resonance structure \cite{Wisniacki11}. 
We are, therefore, led to apply the RADT theory to the strong inter-mode interaction cases in our system. 
 
In the RADT theory for multi-dimensional systems \cite{Almeida84,Lichtenberg83}, 
an effective Hamiltonian describing the motion near nonlinear resonances can be derived by means of the secular perturbation theory. 
In a two-dimensional system, the Hamiltonian can be decomposed as
\begin{equation}
H=H_{0}(I_{1},I_{2})+V(I_{1},I_{2},\theta_{1},\theta_{2}),
\label{2DHamiltonian}
\end{equation}
in terms of action-angle variables $\{\theta_i,I_i\}$, where $H_{0}$ is an integrable Hamiltonian and $V$ is a perturbation which may contain nonintegrable terms. 
A resonance condition arises when $p \frac{dH_{0}}{dI_{1}} = q\frac{dH_{0}}{dI_{2}}$ for co-prime positive integers $p$ and $q$. Following the standard secular perturbation theory, we can then derive a pendulum-like effective Hamiltonian near the $p:q$ resonance as
\begin{equation}
H_{p:q} = \frac{(I-I_{p:q})^2}{2M_{p:q}} + V_{p:q} \cos p \theta ,
\label{pendulum}
\end{equation}
where $I=I_{1}$, $\theta=\theta_{1}-\frac{q}{p}\theta_{2}$, $M_{p:q}^{-1}=\left. ({d^2 H_{0}}/{dI^2})\right|_{I=I_{p:q}}$ with $I_{p:q}$ the action at the resonance (see Ref.\ \cite{Supplement} for derivation details).
The amplitude $V_{p:q}$ characterizes the coupling strength between eigenstates of the integrable Hamiltonian $H_0$. 
The effective Hamiltonian results in a chain-like structure of $p$ islands in the phase space. 

Equation (\ref{pendulum}) then suggests a {\em selection rule} that 
the eigenstate of the unperturbed Hamiltonian of a quantum number $m$ can be strongly coupled to another state of a quantum number $m+ip$ ($i$ integer) \cite{Loeck10,Brodier} with a strength proportional to $V_{p:q}^i$.
In addition, the perturbative amplitude $V_{p:q}$ can be inferred from the phase space structure, or more precisely,
the phase-space area $S_{p:q}$ enclosed by the separatrix associated with the $(p:q)$ nonlinear resonance chain, as indicated in the inset of Fig.\ 3(b). 
Specifically, we find the relation \cite{Supplement}
\begin{equation}
V_{p:q}= \frac{S_{p:q}^2}{256M_{p:q}},
\label{VSrelation}
\end{equation}
indicating that the coupling strength in AC would vary in proportion to $S_{p:q}^2$, which can be varied by changing the degree of cavity deformation. Note that $S_{p:q}$ remains invariant under the canonical transformation \cite{Eltschka05}.

%\section{Selection Rule}

To verify the above selection rule for strong interaction in our system, we need to know the angular mode numbers $m$'s of the unperturbed modes associated with the interacting quasi-eigenmodes and compare their difference $\Delta m$ with the number of islands $p$ in the related resonance chain structure. 
A quasi-eigenmode can be considered being almost unperturbed when it is far away from the interaction regions with other modes.
We can thus infer the angular mode number $m$ of its corresponding unperturbed mode by inspecting the spatial mode distribution of the quasi-eigenmode far away from the interaction regions
as shown in Fig.\ \ref{Spectrum}(b).  Note that the angular mode number $m$ is just the half of the number of antinodes in the spatial mode distribution.
In each mode group, the angular mode number increases by 1 when we move up in $ka$ by one free spectral range along the diabatic line in Fig.\ 1(a).
The radial mode number $l$, also called the mode order, of the associated unperturbed mode can also be identified by counting the number of anti-nodes in the radial direction.

\begin{table}[b]
\begin{tabular}{|c||c|c||c|}
\hline
$l$ 	& $\Delta m$	 & $p$	 & interaction strength \\
\hline
1 vs.\ 2	& 8	& 8	& strong\\
2 vs.\ 3	& 6	& 6	& strong\\
3 vs.\ 4	& 6	& 6	& strong\\
2 vs.\ 4	&12	& 6	& strong\\
1 vs.\ 3	&14	& NA		& weak\\
1 vs.\ 4	&20	& NA		& weak\\
\hline
\end{tabular}
\caption {Comparison between the observed angular mode number difference $\Delta m$ and the number $p$ of the islands in the related resonance chain structure for various inter-mode interactions. NA stands for `not applicable'.}
\label {Selection Rule}
\end{table}

The result of our examination on the relation between $\Delta m$ and $p$ in several strong- and weak-interaction cases is summarized in Table.\ \ref{Selection Rule}. 
For all of the strong interaction cases in Fig.\ \ref{Spectrum}(a), the angular mode number difference $\Delta m$ is equal to or twice the number $p$ of the islands in the resonance chain structure as projected by the selection rule in the RADT theory. 
For the interactions between $l$=1 and 3 modes and between $l$=1 and 4 modes, however, a 6-island and an 8-island chain structure, respectively, reside between the tori along which the Husimi functions of the interacting quasi-eigenmodes are distributed when they are far from the interaction region. 
For these {\em weak} interactions the selection rule is not satisfied since the $p$ number of 6 (8) is not divisors of the observed $\Delta m$ of 14 (20).

%\section{Coupling Strength}

\begin{figure}
\includegraphics[width=3.4in]{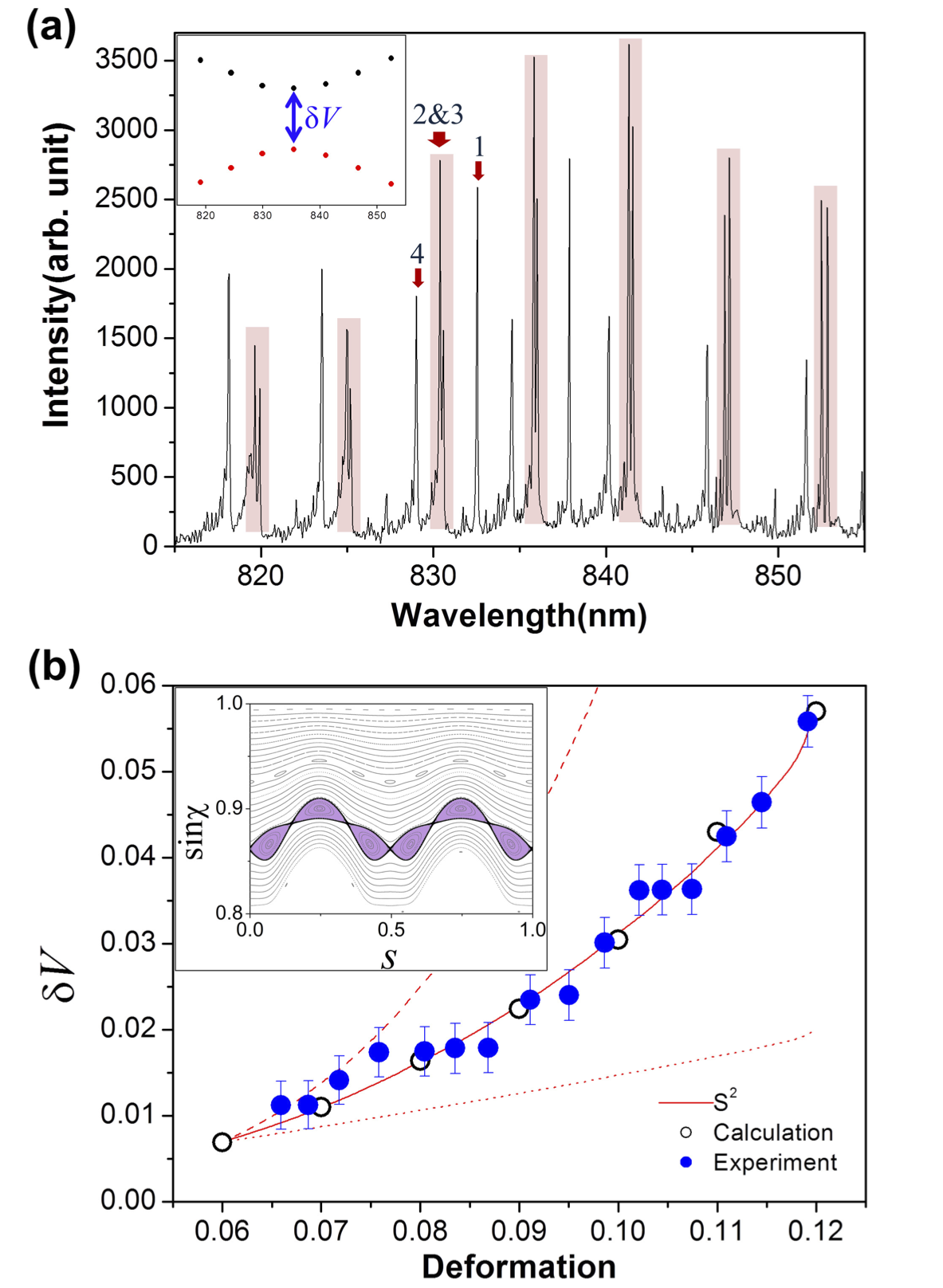}
\caption{(a) Cavity-modified fluorescence spectrum near $\lambda$=835 nm at $\eta = 0.089$. Peaks corresponding to $l$=1, 2, 3 and 4 modes are marked by arrows. (b) $S^2$ (red solid curve) of the 6-island resonance structure is scaled to be compared with the measured AC gaps (blue-filled circles) between $l$=2 and 3 modes near $ka\simeq 114$ and the ones (black-filled circles) from wave calculation for various deformation. 
For comparison, $S$ (red dot curve) and $S^3$ (red dashed curve) curves are also plotted. (inset) PSOS when $\eta=0.08$ and $ka=114$. 
The area $S$ is the area of the shaded region.}
\label{Coupling Strength}
\end{figure}

For experimental observation of RADT, we measured the AC gap of $l$=2 and 3 unperturbed modes for various cavity deformation by using the cavity-modified fluorescence spectroscopy. 
The cavity medium was doped with LDS 821 molecules at a concentration of 0.03 mM/L, covering
a spectral range around $\lambda \simeq$ 832 nm ($ka \simeq$ 114). The cavity deformation $\eta$ was varied from 0.065 to  0.12. A part of the observed spectra is shown in Fig.\ \ref{Coupling Strength}(a), where
among four different mode groups $l$=2 and  3 modes exhibit an AC.

In Fig.\ \ref{Coupling Strength}(b), the observed AC gap $\delta V$ (blue-filled circles) -- defined in the inset of  Fig.\ \ref{Coupling Strength}(a) -- is plotted in the unit of the size parameter as a function of the cavity deformation $\eta$. 
The decay rates of $l$=2 and 3 modes, expected to be less than 1 GHz, are negligible compared to the gap size, which is more than 36 GHz, and thus the gap size is approximately twice the coupling strength between $l$=2 and 3 modes.
The nonlinear resonance involved with this mode interaction is $(p=6, q=1)$ resonance as indicated in the augmented-ray PSOS in Fig.\ \ref{Coupling Strength}(b).
For comparison, the AC gaps (black open circles) from the wave calculation 
and the values of $S_{6:1}^2$ (red solid curve) obtained from the PSOS are also presented in Fig.\ \ref{Coupling Strength}(b). 
Because the proportionality constant $M_{p:q}$ between $V_{p:q}$ and $S_{p:q}^2$ in Eq.\ (3) is not theoretically known for our optical system, it was numerically determined by linear fitting the AC gaps from the wave calculation with $S_{6:1}^2$.
We find that our experimental and numerical results well confirm the $S^2$-dependence of the coupling strength. Interestingly, the AC gaps follow the $S^{2}$ curve even in the moderate perturbation regime with $\eta > 0.10$, where the separatrix shows mild stochasticity.

\begin{figure}
\includegraphics[width=3.4in]{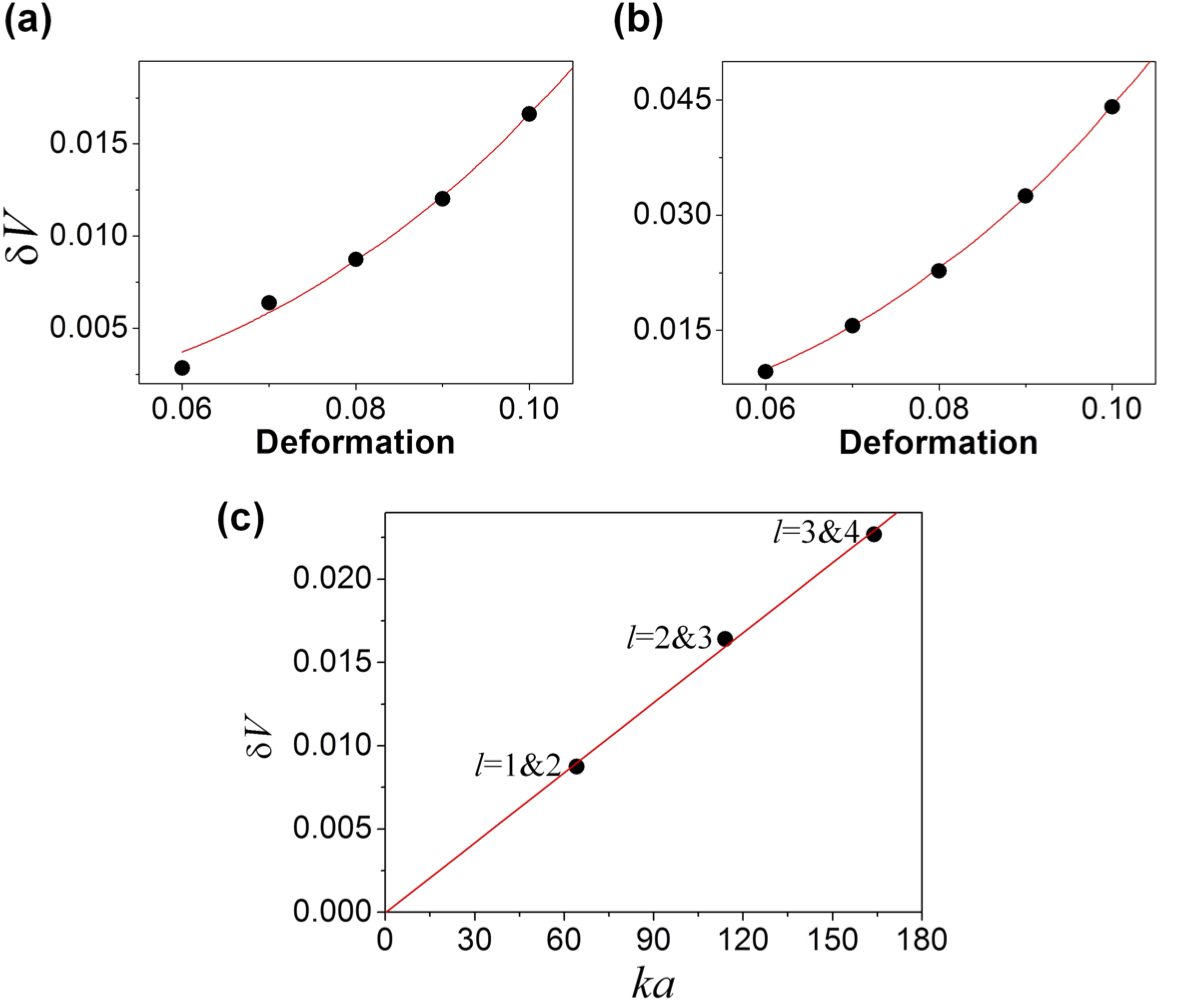}
\caption{(a) $S^2$ of the 6-island resonance structure and the calculated AC gap for the interactions between $l=1$ and 2 modes from $\eta =0.06$ to $\eta = 0.10$. $S^2$ is scaled with respect to the calculated AC gaps in the same manner as in Fig.\ \ref{Coupling Strength}(b). (b) The same for $l=3$  and 4 modes.  
(c) The calculated AC gaps for three different pairs of modes occurring at different $ka$ values when $\eta=0.08$ are all well fit by a common prefactor $\tilde{M}_{6:1}$ of $0.26\pm 0.01$.}
\label{Coupling Calculation}
\end{figure}

For further investigation, we have also performed the wave calculation of the avoided-crossing gaps for $l$=1 and 2 modes near $ka\simeq$ 65 and for $l$=3 and 4 modes near $ka\simeq$ 165. 
Both interactions are related to 6-island resonance chain structures over the deformation range from $\eta$=0.06 to 0.10. 
As we can see in Figs.\ \ref{Coupling Calculation}(a) and \ref{Coupling Calculation}(b), these strong interactions also satisfy the relation $\delta V \propto S^2$. 
Although not presented here, the relation is also satisfied by the strong interaction related to the 8-island resonance chain structure such as $l$=1 and 2 modes near $ka \simeq$ 150.

%\section{prefactor}

As discussed above, the proportionality constant or the prefactor $M_{p:q}$ in Eq.\ (3) was determined by numerically fitting the AC gaps with $S^2$ obtained from the augmented-ray PSOS. Since the PSOS is presented in a dimensionless $(s, \sin \chi)$ phase space, we can rescale Eq.\ (3) as $V_{p:q}=(ka) (\pi^2/64)( \tilde{S}_{p:q}^2 /\tilde{M}_{p:q})$ in $ka$ unit  \cite{Supplement}, where both $\tilde{M}_{p:q}$ and $\tilde{S}_{p:q}$ are dimensionless and $\tilde{S}_{p:q}$ is the separatrix area in the $(s, \sin \chi)$ phase space.
We found that $\tilde{M}_{p:q}$ determined by the fitting depends only on the nonlinear resonance chain that the interacting modes are associated.
For example, when $\eta=0.08$, we observe AC's between $l=1$ and 2 modes at $ka\simeq 65$, between $l=2$ and 3 modes at $ka\simeq 114$  and between $l=3$ and 4 modes at $ka\simeq 165$, respectively. Their AC gaps are well fit by the above rescaled formula with a common $\tilde{M}_{6:1}\simeq 0.26\pm0.01$ as shown in Fig.\ \ref{Coupling Calculation}(c).
This observation elucidates the semiclassical origin of RADT. 
 
%\section{Summary}

In summary, we have experimentally observed the resonance-assisted dynamical tunneling in the inter-mode interactions in an asymmetric microcavity. 
A selection rule for strong interaction mediated by RADT was confirmed and the coupling strength was found to be proportional to the square of the area enclosed by the separatrix associated with the nonlinear resonance chain involved in the interaction.
The present work is a step forward in semiclassical description of the inter-mode interactions in asymmetric microcavities \cite{Takami92, Wiersig06, Lee09}.

We thank S.-B.\ Lee, J. Yang, S.-Y.\ Lee , S.-W.\ Kim and J.-B.\ Shim for helpful discussions. 
This work was supported by the Korea Research Foundation (Grant No.\ WCU-R32-10045).


\begin{thebibliography}{99}

\bibitem{Davis81}
M.\ J.\ Davis and E.\ J.\ Heller, J.\ Chem.\ Phys.\ {\bf 75}, 246 (1981).

%\bibitem{Creagh98}
%S.\ Creagh  in {\em Tunneling in Complex Systems}, edited by S.\ Tomsovic (World Scientific, Singapore, 1998), p. 1.

\bibitem{Lin90}
W.\ A.\ Lin and L.\ E.\ Ballentine, Phys.\ Rev.\ Lett.\ {\bf 65}, 2927
(1990).

\bibitem{Bohigas93}
O.\ Bohigas, S.\ Tomsovic, and D.\ Ullmo, Phys.\ Rep.\ {\bf 223},
43 (1993).

\bibitem{Tomsovic94}
S.\ Tomsovic and D.\ Ullmo, Phys.\ Rev.\ E {\bf 50}, 145 (1994).

\bibitem{Steck01}
D.\ A.\ Steck, W.\ H.\ Oskay and M.\ G.\ Raizen, Science {\bf 293}, 274 (2001).

\bibitem{Hensinger01}
W.\ K.\ Hensinger, H.\ H\"affner, A. Browaeys, N.\ R.\ Heckenberg, K.\ Helmerson, C.\ McKenzie, G.\ J.\ Milburn, W.\ D.\ Phillips, S.\ L.\ Rolston, H.\ Rubinsztein-Dunlop and B.\ Upcroft, Nature {\bf 412}, 52 (2001).

\bibitem{Dembowski00}
C.\ Dembowski, H.-D.\ Gr\"af, A.\ Heine, R.\ Hofferbert, H.\ Rehfeld and A.\ Richter, Phys.\ Rev.\ Lett. {\bf 84}, 867 (2000).

\bibitem{Brodier}
O.\ Brodier, P.\ Schlagheck, and D.\ Ullmo, Phys.\ Rev.\ Lett.\
{\bf 87}, 064101 (2001); Ann.\ Phys.\ (N.Y.) {\bf 300}, 88 (2002).

\bibitem{Almeida84}
A.\ M.\ Ozorio de Almeida, J.\ Phys.\ Chem.\ {\bf 88}, 6139 (1984).

\bibitem{Mouchet06}
A.\ Mouchet, C.\ Eltschka and P.\ Schlagheck, Phys.\ Rev.\ E {\bf 74}, 026211 (2006).

\bibitem{Wimberger06}
S.\ Wimberger, P.\ Schlagheck, C.\ Eltschka and A.\ Buchleitner, Phys.\ Rev.\ Lett.\ {\bf 97}, 043001 (2006).

\bibitem{Sheinman06}
M.\ Sheinman, S.\ Fishman, I.\ Guarneri and L.\ Rebuzzini, Phys.\ Rev.\ A {\bf 73}, 052110 (2006).

\bibitem{Keshavamurthy05}
S.\ Keshavamurthy, J.\ Chem.\ Phys.\ {\bf 122}, 114109 (2005).

\bibitem{Keshavamurthy05r}
S.\ Keshavamurthy, Phys.\ Rev.\ E {\bf 72}, 045203(R) (2005).

\bibitem{Eltschka05}
C.\ Eltschka and P.\ Schlagheck, Phys.\ Rev.\ Lett.\ {\bf 94}, 014101 (2005).

\bibitem{Loeck10}
S.\ L\"ock, A.\ B\"acker, R.\ Ketzmerick and P.\ Schlagheck, Phys.\ Rev.\ Lett.\ {\bf 104} 114101 (2010).

\bibitem{Wisniacki11}
D.\ A.\ Wisniacki, M.\ Saraceno, F.\ J.\ Arranz, R.\ M.\ Benito and F.\ Borondo, Phys.\ Rev.\ E {\bf 84} 026206 (2011).

\bibitem{Yang06}
J.\ Yang, S.\ Moon, S.-B.\ Lee, J.-H.\ Lee and K.\ An, Rev.\ Sci.\ Instrum. {\bf 77}, 083103
(2006).

\bibitem{Moon08}
S.\ Moon, J.\ Yang, S.-B.\ Lee, J.-B.\ Shim, S.-W.\ Kim, J.-H.\ Lee and K.\ An, Opt.\ Express {\bf 16}, 11007 (2008).

\bibitem{Wiersig03}
J.\ Wiersig, J.\ Opt.\ A: Pure\ Appl.\ Opt. {\bf 5}, 53 (2003).

\bibitem{Lee09}
S.-B.\ Lee, J.\ Yang, S.\ Moon, S.-Y.\ Lee, J.-B.\ Shim, S.-W.\ Kim, J.-H.\ Lee and K.\ An, Phys.\ Rev.\ A {\bf 80}, 011802(R) (2009).

%\bibitem{Hentschel03}
%M.\ Hentschel {\em et al.}, Europhys.\ Lett.\ {\bf 62}, 636 (2003).

\bibitem{Unterhinninghofen08}
J.\ Unterhinninghofen, J.\ Wiersig and M.\ Hentschel, Phys.\ Rev.\ E {\bf 78}, 016201 (2008).

\bibitem{Lichtenberg83}
A.\ J.\ Lichtenberg and M.\ A.\ Liebermann, {\em Regular and Stochastic Motion} (Springer, New York, 1983).

\bibitem{Supplement}
See Supplementary Material for recapitulation of the resonance-assisted dynamical tunneling theory in a two-dimensional system and the derivation of Eq.\ (3) and its dimensionless form.

\bibitem{Takami92}
T.\ Takami, Phys.\ Rev.\ Lett.\ {\bf 68}, 3371 (1992).

\bibitem{Wiersig06}
J.\ Wiersig, Phys.\ Rev.\ Lett.\ {\bf 97}, 253901 (2006).

\end{thebibliography}
\end{document}